\documentclass[prd,floats,twocolumn,showpacs,preprintnumbers,showkeys,superscriptaddress]{revtex4-1}

\usepackage{graphicx}

\usepackage{color}

\usepackage{amsfonts}
\usepackage{amssymb}
\usepackage{amsmath}

\newcommand{\figwidth}{\columnwidth}

\newcommand{\non}{\nonumber}
\newcommand{\fm}{\,{\rm fm}}
\newcommand{\mev}{\,{\rm MeV}}
\newcommand{\gev}{\,{\rm GeV}}

\newcommand{\QQ}{{\cal Q}}

\newcommand{\m}{\phantom{-}}

\newcommand{\GE}{G_E}
\newcommand{\GM}{G_M}
\newcommand{\tel}{\tau_{\rm el}}
\newcommand{\muN}{\,\boldmath{\mu}_\mathrm{N}}

\begin{document}

\preprint{
\vbox{
\hbox{ADP-14-22/T880}
}}

\title{Dispersive estimate of the electromagnetic charge symmetry violation\\ in the octet baryon masses}

\author{F.~B.~Erben}
\address{CSSM \& CoEPP, Department of Physics, University of Adelaide, Adelaide SA 5005, Australia}
\address{Institut f\"ur Kernphysik, Becher-Weg 45, University of Mainz,
D-55099 Mainz, Germany}
\author{P.~E.~Shanahan}
\author{A.~W.~Thomas}
\author{R.~D.~Young}

\address{CSSM \& CoEPP, Department of Physics, University of Adelaide, Adelaide SA 5005, Australia}

\begin{abstract}
  We explore the electromagnetic contribution to the charge symmetry
  breaking in the octet baryon masses using a subtracted dispersion relation
  based on the Cottingham formula. For the proton--neutron mass splitting
  we report a minor revision of the recent analysis of Walker-Loud,
  Carlson and Miller. For the electromagnetic structure of the
  hyperons we constrain our analysis, where possible, by a combination
  of lattice QCD and SU(3) symmetry breaking estimates. The results for
  the baryon mass splittings are found to be compatible with recent
  lattice QCD+QED determinations. The uncertainties in the dispersive
  analysis are dominated by the lack of knowledge of the hyperon
  inelastic structure.
\end{abstract}



\maketitle

\section{Introduction}
\label{sec:intro}
A vast array of nuclear and hadronic physics processes are almost
invariant under charge symmetry
\cite{Miller:2006tv,Londergan:2009kj}. As a result, the assumption of
good charge symmetry has been widely applied in nuclear and strong
interaction studies.  With the description of strong interaction
phenomena in terms of the fundamental theory of quantum chromodynamics
(QCD) progressing into the precision era, it is now essential to
further quantify the degree to which charge symmetry is violated ---
see for example the search for new physics in $\beta$ decays
\cite{Gonzalez-Alonso:2013ura}.
Charge symmetry violation (CSV) is driven by two sources, that arising
from the inequality of the light-quark masses ($m_u\ne m_d$), which we
will refer to as the strong component, and that arising from the
electromagnetic interaction.

The prime example of charge symmetry violation (CSV) is the observed
$\sim 0.1\%$ difference in the masses of the proton and
neutron. Calculations in lattice QCD have recently made significant
advances in the determination of the strong component of this mass
difference
\cite{Beane:2006fk,Blum:2010ym,deDivitiis:2011eh,Horsley:2012fw,Shanahan:2012wa,Borsanyi:2013lga}. In
parallel, the theoretical description of the electromagnetic
contribution has been improved by the work of Walker-Loud, Carlson \&
Miller (WCM) \cite{WalkerLoud:2012bg} using a new formulation of the
Cottingham formula \cite{Cottingham:1963zz}. Lattice QCD+QED
\cite{Blum:2010ym,Borsanyi:2013lga,Horsley:2013qka} is also making
progress in the direct calculation of the electromagnetic
contribution.

The principal focus of the present work is the extension of the WCM
dispersive analysis to investigate the electromagnetic contribution to
the mass splittings of the $\Sigma$ and $\Xi$ baryons. The theoretical
inputs required for the dispersion integral are described in terms of
the electromagnetic structure, for which very little is known
phenomenologically for the hyperons.
The results presented here utilise input from lattice QCD, where
available, with conservative estimates of the magnitude
of SU(3) breaking effects applied elsewhere.

In his seminal work \cite{Cottingham:1963zz}, Cottingham showed that
the electromagnetic self-energies of the nucleons can be computed in
terms of the imaginary part of the forward Compton amplitude, which is
measurable in inclusive electron--nucleon scattering
experiments. Using the Cottingham result, the long-standing accepted
value for the electromagnetic contribution to the proton-neutron mass
splitting was $\delta M_{p-n}^\gamma=0.76\pm0.30\mev$
\cite{Gasser:1974wd,Gasser:1982ap}.  The recent work of WCM has
challenged this result by demonstrating that the application of the
Cottingham formula with two different Lorentz decompositions of the
Compton scattering tensor leads to incompatible results
\cite{WalkerLoud:2012bg}. By using a subtracted dispersive analysis,
WCM demonstrated that this ambiguity can be removed. The revised value
of the dispersive estimate of the electromagnetic mass splitting was
reported to be $\delta M_{p-n}^\gamma=1.30\pm 0.47\mev$
\cite{WalkerLoud:2012bg}. An extension of the WCM formalism
\cite{Thomas:2014dxa} which incorporates quark-mass dependence and
finite volume effects, combined with the lattice simulation results of
Ref.~\cite{Blum:2010ym}, provides an improved constraint on the
dispersion integral $\delta M_{p-n}^\gamma=1.04\pm 0.11\mev$.

%
\section{Electromagnetic self-energy}
\label{sec:pheno}

As described by WCM, the use of a subtracted dispersion relation for
the determination of the electromagnetic self-energy of a baryon $B$
leads to the natural separation of contributions given by
\begin{equation}
\delta M_B^\gamma=\delta M_B^{\rm el}+\delta M_B^{\rm inel}+\delta M_B^{\rm sub}+\delta\tilde{M}_B^{\rm ct}\,.
\label{eq:dMgamma}
\end{equation}
In the following subsections, each of these contributions is examined
in the light of our current understanding of nucleon and hyperon
structure.
\subsection{Elastic}
%
The elastic contribution to the self-energy is given by
\begin{align}
\delta M_B^{\rm el}&=
\frac{\alpha}{\pi}\int_0^{\Lambda_0} dQ\bigg[\frac32\GM^2\frac{\sqrt{\tel}}{\tel+1}\non\\
&\ +(\GE^2-2\tel\GM^2)\frac{(1+\tel)^{3/2}-\tel^{3/2}-\tfrac32\sqrt{\tel}}{\tel+1}\bigg],
\label{eq:elastic}
\end{align}
with $\tel=Q^2/(4M_B^2)$. $\GE$ and $\GM$ represent the electric and
magnetic Sachs form factors of the corresponding baryon. For the
proton and neutron, these are rather well-known empirically and we
make use of the Kelly parameterisation \cite{Kelly:2004hm} of
experimental results. The upper limit of integration, $\Lambda_0$,
denotes the scale at which perturbative evolution becomes reliable. We
follow WCM by reporting central estimates using $\Lambda_0^2=2\gev^2$,
and uncertainties calculated by allowing for variation over the range
$1.5<\Lambda_0^2<2.5\gev^2$ \cite{WalkerLoud:2012bg}.

For the hyperons, we use lattice-QCD-based results from the
CSSM/QCDSF/UKQCD Collaborations. The lattice study of
Refs.~\cite{Shanahan:2014uka,Shanahan:2014cga} presents results for
the electromagnetic form factors of all outer-ring octet baryons at a
range of discrete values of the momentum transfer, $Q^2$. The analysis
includes finite-volume corrections and a chiral extrapolation to the
physical pseudoscalar masses. In addition, simple parameterizations of
the $Q^2$-dependence of the form factors are given at the physical
point. It is these parameterizations which we use here.

It was found in Ref.~\cite{Shanahan:2014cga}, for the electric form
factors, that standard dipole parameterizations of the
$Q^2$-dependence of $G_E$ perform poorly. Here, for the charged
baryons, we use the more general fits presented in that work,
\begin{equation}
G_{E,\textrm{fit}}^B(Q^2)=\frac{G^B_E(Q^2=0)}{1+c_1Q^2+c_2Q^4+c_3Q^6}.
\end{equation}
For the neutral cascade baryon form factor, where the charge
$G_E^{\Xi^0}(Q^2=0)=0$, we use the same form, fit to the individual
quark-sector contributions to the form factor. The total form factor
is then deduced as
\begin{align}
\GE^{\Xi^{0/-}}   &=\QQ_{u/d}G_{E,\textrm{fit}}^{\Xi^0,u}(Q^2)+2\QQ_s G_{E,\textrm{fit}}^{\Xi^0,s}(Q^2)\,,
\end{align}
with $\QQ_{u,d,s}$ the charges of the respective quarks. For
consistency this same process is followed for the $\Xi^-$.

Similarly, we take parameterizations of the hyperon magnetic form
factors from Ref.~\cite{Shanahan:2014uka}. The function that best
reproduced the lattice simulation results is
\begin{equation}
G_{M,\textrm{fit}}^B(Q^2)=\frac{\mu_B}{1+c_1Q^2+c_2Q^4+c_3Q^6},
\end{equation}
where $\mu_B$ denotes the experimental value of the magnetic moment of the baryon $B$~\cite{Beringer:1900zz}.
Here, as in Ref.~\cite{Shanahan:2014uka}, $G_{M}$ has been expressed in units of the
nuclear magneton $\muN\equiv e\hbar/(2M_p)$. Note that in order to use these
expressions in Eq.~(\ref{eq:elastic}) one must multiply them by
a factor $M_B/M_p$.
The elastic contributions to the mass splittings are summarised in
Table~\ref{tab:summ}.

\begin{table*}
\caption{Decomposition of the electromagnetic contributions to the octet baryon mass splittings as defined in Eq.~(\ref{eq:dMgamma}).
  \label{tab:summ}}
\begin{center}
\begin{tabular}{l|lllll|l}
\hline\hline
Baryon                        & $\delta M^{\rm el}$ & $\delta M^{\rm inel}$ & $\delta M^{\rm sub}_{\rm el}$ & $\delta M^{\rm sub}_{\rm inel}$ & $\delta \tilde{M}^{\rm ct}$ & $\delta M^{\gamma}$ \\
\hline
$p-n$                         & $\m 1.401(7)$      & $0.089(42)$         & $-0.635(7)$                 & $0.18(35)$                    & $0.006$                    & $\m 1.04(35)$ \\
$\Sigma^+-\Sigma^-$           & $\m 1.24(7)$       & $0.02(21)$          & $-1.89(10)$                 & $0.6(11)$                     & $0.014(1)$                 & $\m 0.0(11)$ \\
$\Xi^0-\Xi^-$                 & $-0.636(30)$       & $0.42(15)$          & $-0.80(4)$                  & $0.6(11)$                     & $0.008$                    & $-0.4(11)$ \\
\hline\hline
\end{tabular}
\end{center}
\end{table*}

\subsection{Inelastic}
%
The inelastic contribution to the electromagnetic self-energy can be expressed in the form
\begin{align}
\delta M^{\rm inel}_B&=\int_{W^2_0}^\infty dW^2\,\Omega_B^{\rm inel}(W^2),
\end{align}
where
\begin{align}
\Omega^{\rm inel}(W^2)&=\non\\
&\hspace*{-10mm}\frac{\alpha}{\pi}\int_0^{\Lambda_0} dQ
\bigg\{\frac{3F_1(W^2,Q^2)}{4M_B^2}\frac{2\tau^{\tfrac32}-2\tau\sqrt{1+\tau}+\sqrt{\tau}}{\tau}
\non\\
&\hspace{-5mm}\phantom{\bigg\{}
+\frac{F_2(W^2,Q^2)}{(Q^2+W^2-M_B^2)}\left[(1+\tau)^{\tfrac32}-\tau^{\tfrac32}-\tfrac32\sqrt{\tau}\right]\bigg\}
\label{eq:inel}
\end{align}
with $\tau=(W^2+Q^2-M_B^2)^2/(4M_B^2Q^2)$ and $W_0=(M_B+m_\pi)$.
$F_1$ and $F_2$ denote the baryon inelastic structure functions. We
note that the standard derivation of the dispersion integral yields an
integral with respect to $\nu$, the energy transferred to the
target. Here we have transformed the integration variable $\nu\to
W^2$, where $W^2$ is the invariant mass-squared of the hadronic intermediate state, in
order to highlight the distinct resonance structures.

The structure functions $F_1$ and $F_2$ have been measured extensively
for the proton and deuteron. For the low to intermediate $W$ region we
make use of the parameterisations of Christy \& Bosted (CB)
\cite{Christy:2007ve,Bosted:2007xd}.  
As nearly all data points agree with the proton structure function parameterisations to better than 5\%,
we take the conservative estimate of a uniform 5\% uncertainty in
$F_{1,2}^p$. The parameterisation of the deuteron scattering data is
in similar agreement at the 3--5\% level \cite{Christy:2007ve}, with
some data points out to $\sim$10\% disagreement in limited kinematic
domains. Since the neutron structure functions are estimated by
subtracting out the knowledge of the proton, we assign a conservative
10\% uncertainty on the neutron structure functions.

\begin{figure}
\includegraphics[width=\figwidth]{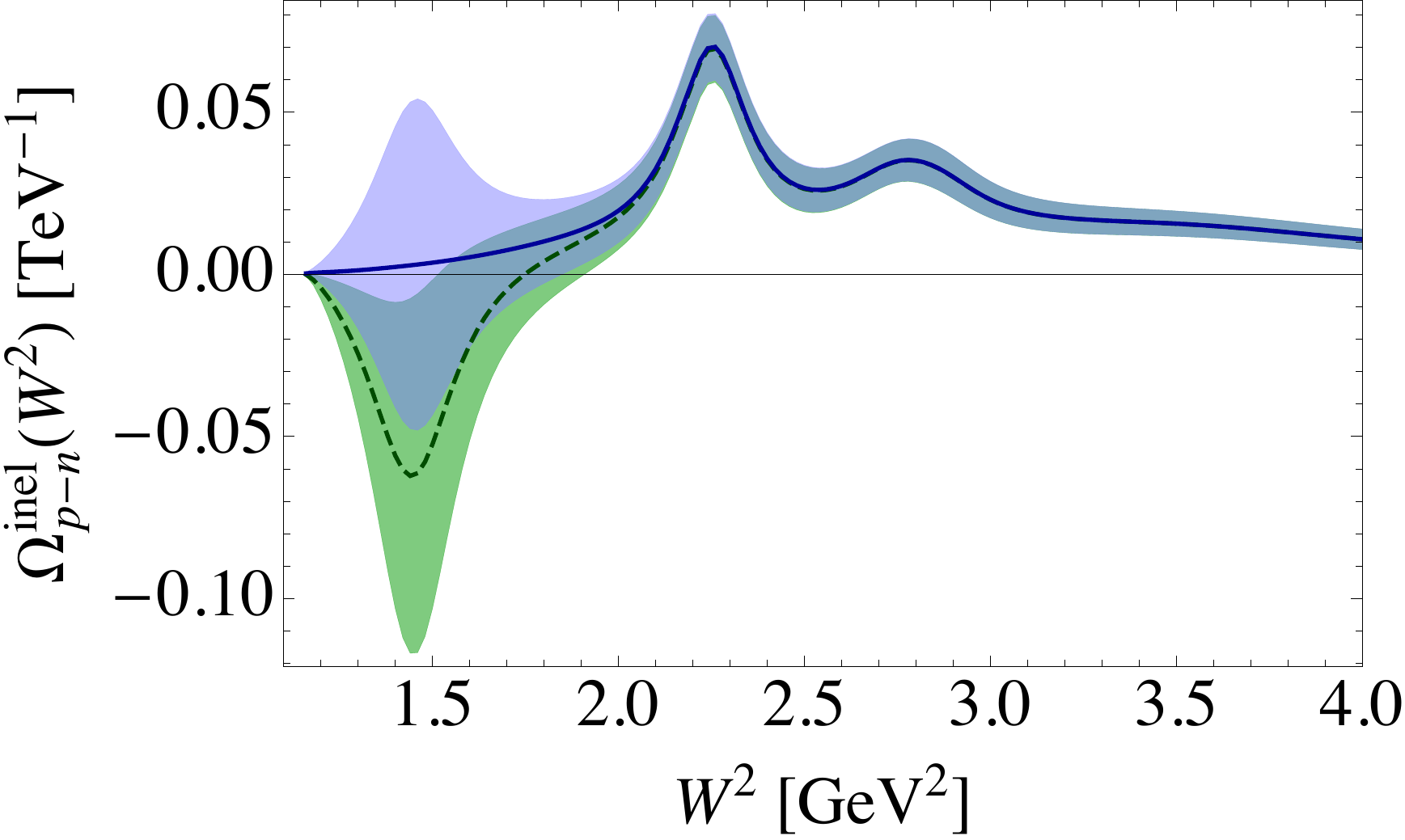}
\caption{The integrand (with respect to $W^2$) of the inelastic dispersion integral contributing to the $p-n$ electromagnetic self-energy (shown for $\mu^2=2\gev^2$). The dotted line shows the result of the direct application of the Bosted-Christy structure functions. The solid line shows the same quantity where the Delta resonance contribution has been forced to be isospin symmetric. In both cases the shaded regions reflect a characteristic uncertainty in the parameterisations of the individual structure functions.
  \label{fig:inel}}
\end{figure}

Figure \ref{fig:inel}
displays the integrand $\Omega_{p-n}^{\rm inel}(W^2)$ contributing to
the proton--neutron mass splitting calculated using the CB parameterisations. Under exact charge symmetry, the
cross sections for $\gamma^*p\to\Delta^+$ and $\gamma^*n\to\Delta^0$
are identical. The central values of the Bosted \& Christy
parameterisation give a violation of this symmetry by about 18\% in
the Delta production rate.  This significant CSV effect is what causes
the large dip structure seen in Fig.~\ref{fig:inel} in the Delta
region.
While we expect some CSV in the Delta region the CB value seem
excessively large. Bearing in mind that such effects are inextricably
linked with the extraction of the photo-neutron cross section for the
deuteron,
in the present analysis
we prefer to take a charge symmetric Delta
production rate as our central value.
To achieve this, we set the Delta parameters of the Bosted-Christy
deuteron fits to match those of the proton results. We attach a 100\%
uncertainty to this artificial modification of the empirical fits.
This modification leads to an appreciable change in the cross
sections only in the difficult-to-constrain low-$Q$ and low-$W$
region.
As a consequence of restoring charge symmetry to the Delta region, the
central value of $\delta M_{p-n}^{\rm inel}$ is increased by just
$0.020\mev$.

For the region $W^2>9\gev^2$ we use the Regge form for the inelastic
structure functions proposed by Capella et al.~\cite{Capella:1994cr},
with the modifications summarised by Sibirtsev et
al.~\cite{Sibirtsev:2010zg}.

In summary, we determine the inelastic contributions to the dispersion
integral for the nucleons to be
\begin{align}
\delta M_p^{\rm inel} &=0.62\pm 0.03 \pm 0.07, \\
\delta M_n^{\rm inel} &=0.53\pm 0.05 \pm 0.05,\\
\delta M_{p-n}^{\rm inel} &=0.089\pm 0.038 \pm 0.019,
\end{align}
where the first error is that from the uncertainty associated with the
structure functions and the second is from the range of $\Lambda_0^2$.

Very little is known experimentally about the hyperon structure
functions. There are some older studies based on the MIT bag model
\cite{Boros:1999tb}, while recent lattice QCD studies have provided
insight into the partonic structure of the octet baryons
\cite{Horsley:2010th,Cloet:2012db}.
These simulations offer some
guidance as to the size of SU(3) breaking effects in the inelastic
structure functions. Based on the results of a recent chiral
extrapolation \cite{Shanahan:2013vla}, we report estimates for the
ratios of the quark momentum fractions at the physical quark masses:
\begin{align}
R_u^\Sigma\equiv\frac{\langle x\rangle_u^\Sigma}{\langle x\rangle_u^p}&
=1.2(1),\quad
R_d^\Sigma\equiv\frac{\langle x\rangle_s^\Sigma}{\langle x\rangle_d^p}
=1.5(1),
\label{eq:R1}\\
R_u^\Xi\equiv \frac{\langle x\rangle_s^\Xi}{\langle x\rangle_u^p}  &
=1.19(4),\quad
 R_d^\Xi\equiv \frac{\langle x\rangle_u^\Xi}{\langle x\rangle_d^p} 
=1.4(2).
\label{eq:R2}
\end{align}
While the partonic interpretation is not generally applicable at the
low-$Q^2$ values of relevance to the integral of Eq.~(\ref{eq:inel}),
we will adopt the flavour separation to enable us to use these lattice
estimates, Eqs.~(\ref{eq:R1}) \& (\ref{eq:R2}), to guide the
significance of the SU(3) breaking. We write the up or down
contributions to the nucleon structure functions in terms of the
proton and neutron structure functions as
\begin{equation}
F^{N,u}=\frac{9}{15}\left(4F^p-F^n\right)
,\quad F^{N,d}=\frac{9}{15}\left(4F^n-F^p\right).
\end{equation}
Here we have assumed partonic charge symmetry, i.e., $F^{N,u}\equiv
F^{p,u}=F^{n,d}$ and $F^{N,d}\equiv F^{p,d}=F^{n,u}$. To estimate the inelastic
self-energies of Eq.~(\ref{eq:inel}) we use structure functions
that are scaled by the lattice estimates
\begin{align}
F^{\Sigma,u}\simeq\frac{\langle x\rangle_u^\Sigma}{\langle x\rangle_u^p}F^{N,u},\quad
F^{\Sigma,s}\simeq\frac{\langle x\rangle_s^\Sigma}{\langle x\rangle_d^p}F^{N,d},\\
F^{\Xi,s}\simeq\frac{\langle x\rangle_s^\Xi}{\langle x\rangle_u^p}F^{N,u},\quad
F^{\Xi,u}\simeq\frac{\langle x\rangle_u^\Xi}{\langle x\rangle_d^p}F^{N,d}.
\end{align}
We caution that the resonance structures in the hyperons
are markedly different from those in the nucleons. Nevertheless,
the success of duality in the case of the nucleon
\cite{Melnitchouk:2005zr} suggests that such $W^2$-integrated quantities may be
reasonably estimated by this simple SU(3) scaling. This assumption
could be improved upon with a more thorough analysis of the flavour
separation in the low-$Q^2$ region, such as that explored in
Refs.~\cite{Rislow:2010vi,Gorchtein:2011mz,Hall:2013hta}. Given the relatively
small magnitude of $\delta M^{\rm inel}$, such an improvement is not warranted
in the present calculation.

Under the assumptions stated previously, we can estimate the hyperon
inelastic integrals in terms of the corresponding nucleon
results. Explicitly,
\begin{align}
\delta M_{\Sigma^{+}-\Sigma^-}^{\rm inel}&=\left(\QQ_{u}^2-\QQ_d^2\right)\frac{9}{15}R_u^\Sigma\left(4\delta M_p^{\rm inel}-\delta M_n^{\rm inel}\right),\\
\delta M_{\Xi^0-\Xi^-}^{\rm inel}&=\left(\QQ_{u}^2-\QQ_d^2\right)\frac{4}{15}R_d^\Xi\left(4\delta M_n^{\rm inel}-\delta M_p^{\rm inel}\right).
\end{align}
For a conservative estimate of the uncertainties, we include an
uncertainty on the lattice momentum fraction ratios ($R_q^B$) that allows for a
100\% variation of the amount of SU(3) violation (i.e.~$R_q^B-1$). The final
results for the hyperon inelastic integrals are summarised in
Table~\ref{tab:summ}.

\subsection{Subtraction}
%
Using the subtracted dispersion formalism of WCM, one is left with a
dependence of the self-energy on the real part of the forward Compton amplitude evaluated
at $\nu=0$ \cite{WalkerLoud:2012bg}
\begin{equation}
\delta M^{\rm sub}_B=-\frac{3\alpha}{16\pi M_B}\int_0^{\Lambda_0^2}dQ^2\,T_1^B(0,Q^2),
\end{equation}
(see Ref.~\cite{WalkerLoud:2012bg} for the Lorentz decomposition of
the Compton amplitude). The amplitude $T_1(0,Q^2)$ has received
considerable attention recently
\cite{Carlson:2011zd,Hill:2011wy,Birse:2012eb} in relation to the
proton radius puzzle \cite{Pohl:2010zza,Pohl:2013yb}. Knowledge of the
momentum dependence of $T_1$ can be expressed as
\begin{equation}
T_1^B(0,Q^2)=2G_M^2(Q^2)-2F_D^2(Q^2)+Q^2\frac{2M_B}{\alpha}\beta_M^B F_\beta(Q^2),
\label{eq:T1sub}
\end{equation}
where $F_D$ denotes the elastic Dirac form factor. The first two terms
in this expression can naturally be described as the elastic
contribution. This contribution to the self-energy,
\begin{equation}
\delta M_{\rm el}^{\rm sub}=-\frac{3\alpha}{16\pi M}\int_0^{\Lambda_0^2} dQ^2\left[2G_M^2(Q^2)-2F_D^2(Q^2)\right]
\end{equation}
is readily evaluated using the form factors described above.  The
results are displayed in Table~\ref{tab:summ}.

The final term in Eq.~(\ref{eq:T1sub}) describes an inelastic
component, which, as in the calculation of WCM, constitutes the dominant
uncertainty in the calculation. In a small-$Q^2$ expansion of this
component the leading term is given by the magnetic polarisability
\cite{Bernabeu:1976jq}.  A recent phenomenological analysis of the
nucleon magnetic polarizabilities has reported
\cite{Griesshammer:2012we}
\begin{align}
\beta_M^p&=(3.1\pm0.8)\times 10^{-4}\fm^3,\\
\beta_M^n&=(4.1\pm2.0)\times 10^{-4}\fm^3,\\
\beta_M^{p-n}&=(-1.0\pm2.0)\times 10^{-4}\fm^3.
\end{align}

Beyond leading order, the $Q^2$ dependence of the inelastic
contribution is encoded in the form factor $F_\beta(Q^2)$.
Using chiral perturbation theory, Birse and McGovern
\cite{Birse:2012eb} have recently estimated that the small $Q^2$
behaviour of $F_\beta$ for the proton may be described as
\begin{equation}
F_\beta=1+\frac{Q^2}{M_\beta^2}+{\cal O}(Q^4)
\label{eq:radius}
\end{equation}
with a mass scale
\begin{equation}
M_\beta=460\pm100\pm40\mev.
\end{equation}

At large $Q^2$, $T_1$ must fall like $1/Q^2$, as determined by the
operator product expansion \cite{Collins:1978hi}. 
Collins has determined the coefficient of this dominant contribution at large $Q^2$
\cite{Collins:1978hi}:
\begin{widetext}
\begin{equation}
T_1^B(0,Q^2)\stackrel{Q^2\to\infty}{=}\frac{1}{Q^2}\left\{4\kappa M_B^2-4\sum_q\left(\kappa+\QQ_q^2\right)M_B\sigma_q^B+{\cal O}\left[\frac{1}{\log Q^2}\right]\right\}\,,
\end{equation}
%
where to lowest order in the strong coupling $\kappa=N_f/(33-2N_f)$,
the sum is over $N_f$ active flavours of quark $q$ and $\sigma_q^B$
denotes the sigma term for quark flavour $q$ in baryon $B$.
The flavour-dependent sigma terms, including charge symmetry violating
effects, have been studied in recent lattice QCD analyses
\cite{Horsley:2012fw,Shanahan:2012wa}.  The explicit flavour
decomposition, based on the work reported in
Refs.~\cite{Shanahan:2012wa,Shanahan:2013cd,Shanahan:2012wh}, is
displayed in Table~\ref{tab:sigma}.

To leading order in the isospin splittings, and still to first order in
$\alpha$ (i.e., this term amounts to an ${\cal O}(\alpha(m_d-m_u))$
effect), only the isovector contribution is required and the large-$Q^2$ scaling can be written as
\begin{equation}
T_1^{\Delta B}(0,Q^2)\stackrel{Q^2\to\infty}{=}\frac{1}{Q^2}
\left\{-4M_{\bar{B}}\left(\QQ_u^2\frac{m_u}{\bar{m}}-\QQ_d^2\frac{m_d}{\bar{m}}\right)\left(\sigma_u^{\bar{B}}-\sigma_d^{\bar{B}}\right)+{\cal O}\left[\frac{1}{\log Q^2}\right]\right\}\,,
\label{eq:OPEIV}
\end{equation}
\end{widetext}
where we have introduced the isospin-averaged baryon masses $M_{\bar
  B}$ for $\bar{B}=\{N,\Sigma,\Xi\}$ and the light quark masses,
$m_u$, $m_d$ and $\bar{m}=(m_u+m_d)/2$. The isospin-averaged sigma
terms are given by $\sigma_u^N=(\sigma_u^p+\sigma_d^n)/2$,
$\sigma_d^N=(\sigma_d^p+\sigma_u^n)/2$, and similarly for the hyperon
cases. Numerically, $T_1^{\Delta N}(0,Q^2)$ for the nucleon is of the
order $(-2\times 10^{-3}\gev^2)/Q^2$.

\begin{table}
\caption{Flavour break down of light-quark sigma terms (all in MeV). \label{tab:sigma}}
\begin{center}
\begin{tabular}{lcccccc}
\hline\hline
Baryon       & $p$   & $n$   & $\Sigma^+$ & $\Sigma^-$ & $\Xi^0$ & $\Xi^-$ \\
\hline
$\sigma_u^B$ & 18(2) & 14(1) & 13.3(9)    & 3.8(6)     & 7.1(4)  & 1.3(2) \\
$\sigma_d^B$ & 26(3) & 32(3) & 7(1)       & 23(2)      & 2.4(4)  & 12.7(8) \\
\hline\hline
\end{tabular}
\end{center}
\end{table}

Given that the elastic form factors of the nucleon drop off at least
as fast as $1/Q^2$, the elastic component in Eq.~(\ref{eq:T1sub}) is
irrelevant to the large-$Q^2$ behaviour of $T_1(0,Q^2)$. 
Previous authors have advocated
approximating $F_\beta$ in the small \cite{Birse:2012eb} to
intermediate \cite{WalkerLoud:2012bg} $Q^2$ region by a dipole form
\begin{equation}
F_\beta(Q^2)=\left(\frac{1}{1+Q^2/(2M_\beta^2)}\right)^2.
\end{equation}
While these authors have not suggested extending this form to
asymptotically large $Q^2$, we note that this form does not give a
consistent description of the leading $1/Q^2$ behaviour described
above.  Taking the central value for the nucleon isovector
polarisability, $\beta_M^{p-n}\sim -1\times 10^{-4}\fm^3$, in
Eq.~(\ref{eq:T1sub}) with this dipole form and hadronic mass scale
leads to a scaling behaviour $T_1^{\Delta N}(0,Q^2)\sim
-0.8\gev^2/Q^2$. This is a factor of $\sim 400$ larger than predicted
by the operator product expansion.

To smoothly connect the small-$Q^2$ and asymptotic domains, we therefore
suggest a model for the inelastic part of Eq.~(\ref{eq:T1sub}):
\begin{align}
Q^2\frac{2M_{\bar{B}}}{\alpha}\beta_M^{\Delta B} F_\beta^{\Delta B}(Q^2)&\non\\
&\hspace*{-25mm}=\frac{Q^22M_{\bar{B}}\beta_M^{\Delta B}/\alpha +Q^4C_{\Delta B}/(3M_\beta^2)^3}{(1+Q^2/(3M_\beta^2))^3}\,,
\label{eq:T1mod}
\end{align}
where $C_{\Delta B}$ is defined to describe exactly the dominant
contribution to the operator product expansion dependence computed in
Eq.~(\ref{eq:OPEIV}). We note that because the coefficient $C_{\Delta
  B}$ is so small compared to the hadronic scale, it has no
influence on the small-$Q^2$ expansion characterised by the mass scale
$M_\beta$ in Eq.~(\ref{eq:radius}).

Evaluation of the inelastic part of the subtraction term for the nucleon gives
\begin{equation}
\delta M_{\rm inel}^{p-n,\rm sub}=0.18\pm 0.35\mev,
\end{equation}
where the uncertainty reflects the limited
knowledge of $\beta_M^{p-n}$ and mass scale $M_\beta$.
The quoted uncertainty range has been estimated by assuming $\beta_M$
and $\log M_\beta$ to be normally distributed.

Polarizabilities of the hyperons are even less well known than those
of the nucleon. A range of results have been obtained using a variety
of theoretical approaches including chiral effective field theory
\cite{Bernard:1992xi}; soliton models \cite{Gobbi:1995de}; $1/N_C$
expansions \cite{Tanushi:1999rq}; a computational hadronic model
\cite{Aleksejevs:2010zw}; and lattice QCD \cite{Lee:2005dq}. In the
present work we simply take the same value and uncertainty range for
the isovector hyperon polarisabilities as quoted for the nucleon.
The mass scale $M_\beta$ associated with the hyperons has not been
investigated. Since the physics is governed more considerably by the
strange quarks, however, one may anticipate a harder scale than that
for the nucleon. For this reason we take a more conservative range of
mass scales for the hyperons, $M_\beta^{\Sigma,\Xi}=0.7\pm 0.3\gev$.
The resulting contributions to the sum rule are given by
\begin{align}
\delta M_{\rm inel}^{\Sigma^+-\Sigma^-,\rm sub}=0.6\pm 1.1\mev,\\
\delta M_{\rm inel}^{\Xi^0-\Xi^-,\rm sub}=0.6\pm 1.1\mev.
\end{align}
As for the nucleon case, the uncertainties have been propagated
assuming $\beta_M$ and $\log M_\beta$ to be normally distributed.

\subsection{Counter terms}
%
The decompostion of the baryon mass splittings into electromagnetic
and strong components is itself scale dependent. For sufficiently
large $\Lambda_0$, where perturbative QCD is applicable, this scale
dependence is entirely encoded in the operator product expansion
analysis described above.
Although the leading contributions are formally second order for the
charge symmetry violating effects, we include them for completeness.
This leading counterterm evaluates to
\begin{equation}
\delta\tilde{M}_{\Delta B}^{\rm ct}=-\frac{3\alpha}{16\pi M_{\bar B}}C_{\Delta B}\log\left(\frac{\Lambda_0^2}{\Lambda_1^2}\right),
\end{equation}
where, following WCM, we have taken $\Lambda_0=2\gev^2$ and
$\Lambda_1^2=100\gev^2$ for our numerical values, which are summarised
in Table~\ref{tab:summ}.

%
\section{Total}
In summary, our best estimates for the electromagnetic contribution
to the baryon isospin mass splittings are
\begin{align}
\delta M_{p-n}^\gamma & = 1.04\pm 0.35\mev,\\
\delta M_{\Sigma^+-\Sigma^-}^\gamma & = 0.0\pm 1.1\mev,\\
\delta M_{\Xi^0-\Xi^-}^\gamma & = -0.4\pm 1.1\mev.
\end{align}
The value for the isospin breaking in the nucleon sector is compatible
with the analysis by Walker-Loud et al. \cite{WalkerLoud:2012bg}. It
is also in excellent agreement with the dispersion relation
constrained by lattice QCD simulations \cite{Thomas:2014dxa}.

In the hyperon sector, our findings compare
favourably with lattice QCD+QED simulations from the BMW
Collaboration \cite{Borsanyi:2013lga}
\begin{align}
\delta M_{p-n}^\gamma & = 1.59\pm 0.46\mev,\\
\delta M_{\Sigma^+-\Sigma^-}^\gamma & = 0.08\pm 0.36\mev,\\
\delta M_{\Xi^0-\Xi^-}^\gamma & = -1.29\pm 0.17\mev.
\end{align}

\begin{figure}
\includegraphics[width=\columnwidth]{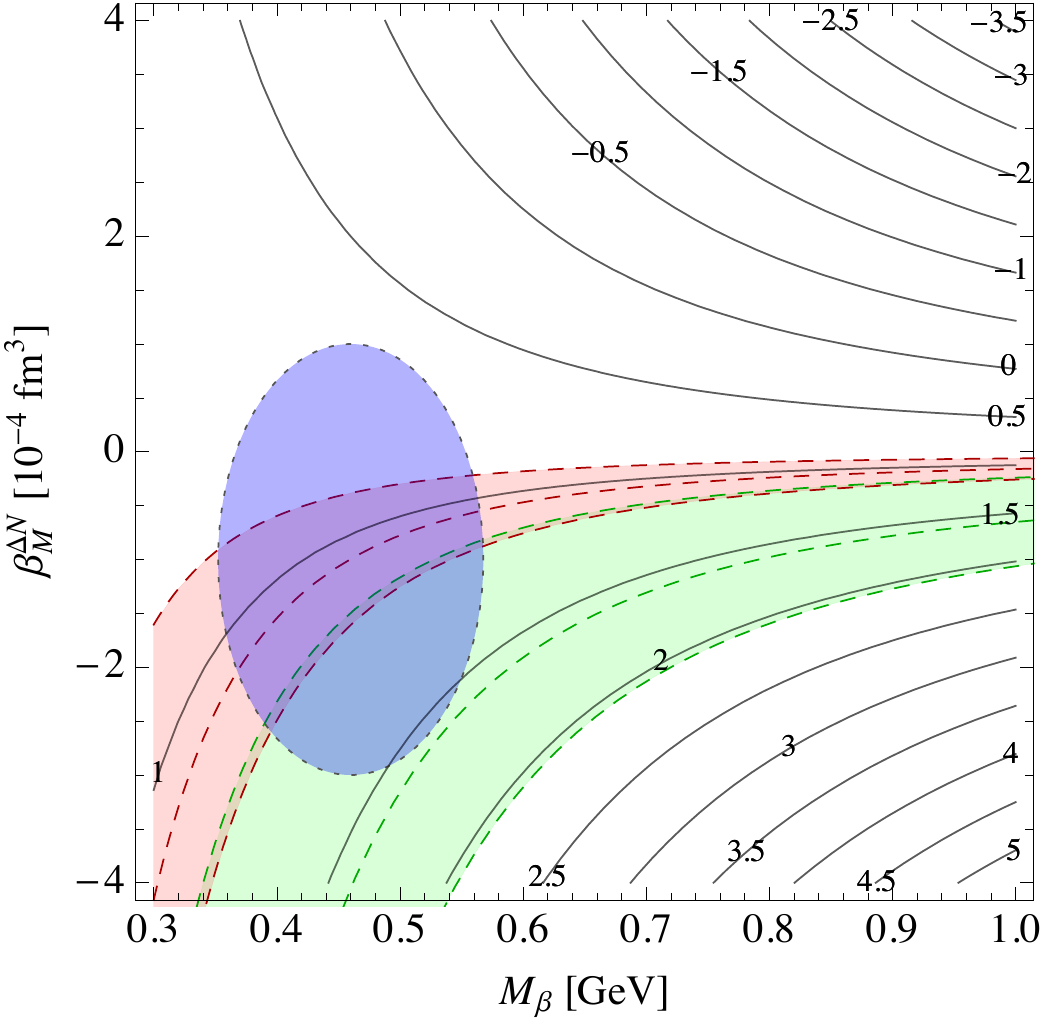}
\caption{The contours depict constant electromagnetic
  self-energy with respect to the dominant driving uncertainties, the
  isovector magnetic polarisability $\beta_M^{p-n}$ and the mass
  parameter $M_\beta$ (see Eq.~(\ref{eq:T1mod})) characterising the mass scale by which the
  corresponding integral is suppressed. The contours are labelled in 
  units of MeV, with the error bar on these lines implied at the level 
  of $\pm 0.04\mev$. The blue ellipse denotes the
  best phenomenological estimates of these parameters as reported in
  Refs.~\cite{Griesshammer:2012we} and \cite{Birse:2012eb},
  respectively. The shaded green band displays the lattice
  calculation of the electromagnetic self energy reported by the BMW
  Collaboration \cite{Borsanyi:2013lga}. The red band shows the lattice-constrained 
  dispersive estimate of $\delta M_{p-n}^\gamma$ reported in Ref.~\cite{Thomas:2014dxa}.
  \label{fig:nucleonPol}}
\end{figure}

As in the work of WCM, the uncertainty of the dispersion integral is
dominated by the lack of knowledge of the inelastic subtraction term.
Here we summarise the intermediate stage of the calculation, computing
all contributions up to this isolated term:
\begin{align}
\delta M_{p-n}^\gamma - \delta M_{\rm inel}^{p-n,\rm sub} & = 0.86\pm 0.04\mev,\label{eq:dMnuclbar}\\
\delta M_{\Sigma^+-\Sigma^-}^\gamma - \delta M_{\rm inel}^{\Sigma^+-\Sigma^-,\rm sub}& = -0.62 \pm 0.24\mev,\\
\delta M_{\Xi^0-\Xi^-}^\gamma -\delta M_{\rm inel}^{\Xi^0-\Xi^-,\rm sub}& = -1.00 \pm 0.16\mev.
\end{align}
With these terms relatively well constrained, the lattice calculation
of the total electromagnetic contribution allows us to explore the
driving uncertainties in the inelastic subtraction
term. Figure~\ref{fig:nucleonPol} displays the dependence of the
nucleon electromagnetic mass splitting on the dominant uncertainties
of the inelastic subtraction term. Compatibility between the
dispersion calculation and lattice is observed. Unfortunately, given
the present central values, it is difficult to improve the
estimates for either $\beta_M$ or $M_\beta$.

\begin{figure}[!t]
\includegraphics[width=0.9\columnwidth]{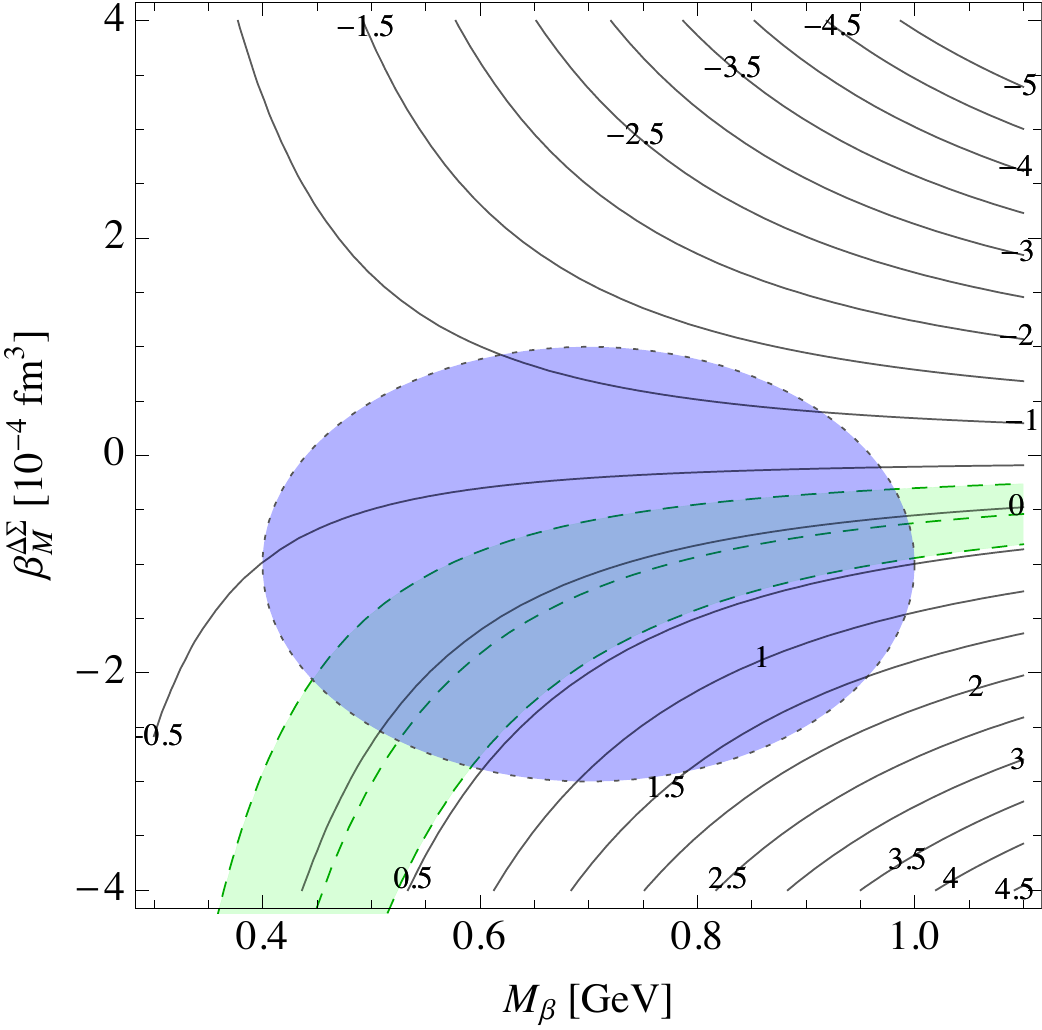}
\includegraphics[width=0.9\columnwidth]{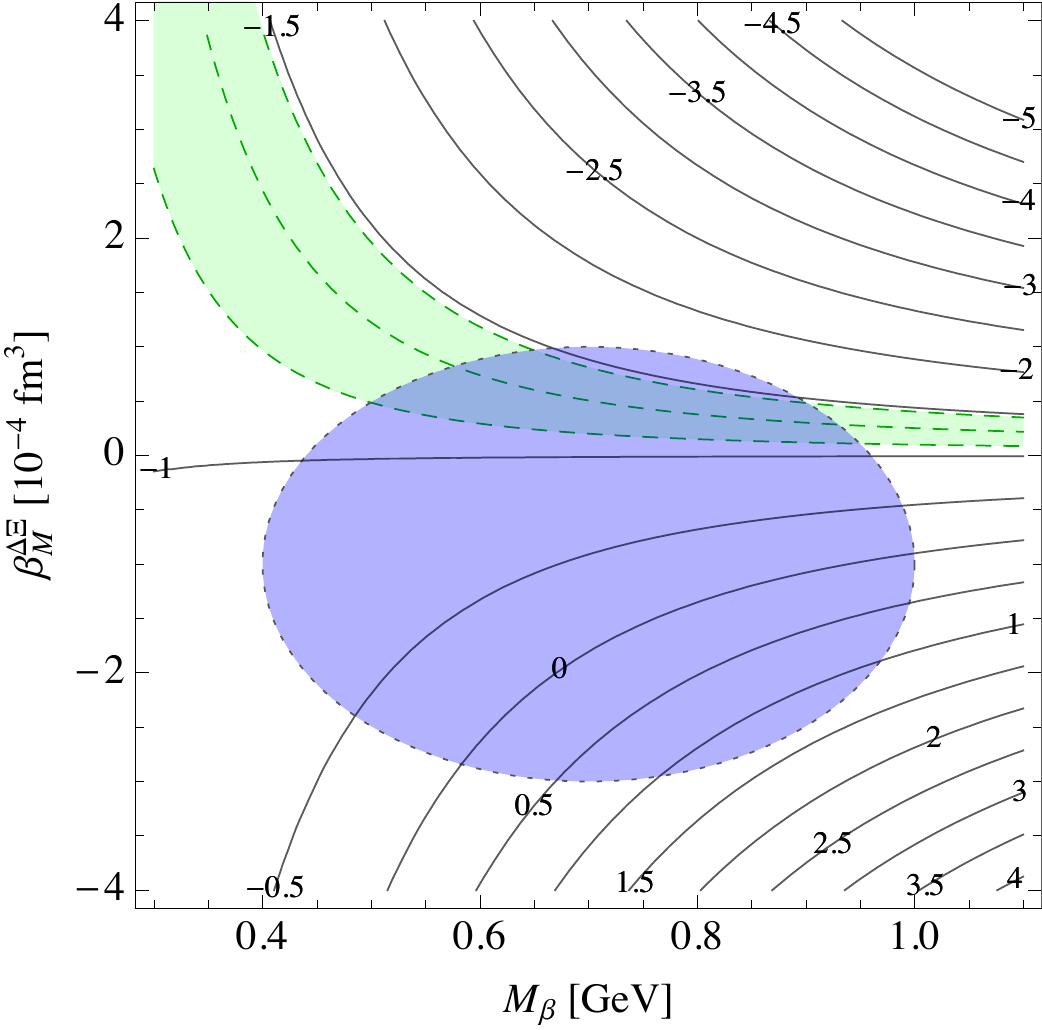}
\caption{Graph is labelled the same as Fig.~\ref{fig:nucleonPol},
  showing the sesitivity of the $\Sigma$ (top panel) and $\Xi$ (lower
  panel) baryon electromagnetic splittings to $\beta_M^{\Delta B}$ and
  $M_\beta$. Uncertainties on the black contours should be interpreted
  as $\pm 0.24\mev$ for $\Sigma$ and $\pm 0.16\mev$ for $\Xi$.
\label{fig:hyperonPol}}
\end{figure}

In Figure~\ref{fig:hyperonPol} we show similar comparison of the
dispersion calculation with the lattice QCD+QED values of the
electromagnetic mass differences. Even with the large range of
$\Lambda_\beta$ considered, it is evident the lattice results can play
some meaningful constraint on the hyperon isovector
polarisabilities. The figures suggest that
$\beta_M^{\Sigma^+-\Sigma^-}$ lies in the range $(-3\to 0)\cdot
10^{-4}\fm^3$ and $\beta_M^{\Xi^0-\Xi^-}$ in the range $(0\to
1.5)\cdot 10^{-4}\fm^3$. If $M_\beta$ turns out to be similarly
soft, as suggested for the nucleon, then less restrictive bounds on
the hyperon polarisabilities would result.

%
\section{Summary}
\label{sec:fin}
We have reported a new analysis of the Cottingham sum rule evaluation
of the electromagnetic contribution to mass differences in the octet
baryon states. We have adapted the recently formulated subtracted
dispersion approach introduced by Walker-Loud et al. to the hyperons,
and implemented some minor updates for the proton-neutron system.
Comparing with this earlier phenomenological work, the minor
differences in the nucleon analysis arise from two sources: i) in this
work, the significant CSV effects in the Delta region realised by the
Bosted-Christy structure functions have been suppressed, this
generates a rather small increase in the self energy; ii) the
inelastic subtraction involving $T_1^{p-n}(0,Q^2)$ is suppressed more
rapidly in this work in order to appropriately match onto the
behaviour dictated by the operator product expansion. This acts to
reduce the size of this term, and consequently lessen the sensitivity
to the poorly-known isovector polarisability.

For the hyperons, the dispersive estimates have significantly larger
uncertainties than for the nucleon, which are dominated by the lack of
knowledge of the hyperon isovector polarisabilities. Comparison with
recent lattice QCD+QED simulations suggests some modest bounds on the
size of the isovector magnetic polarisabilities.
Certainly further theoretical (or experimental) work on this
aspect of hyperon structure would be of interest.

During the completion of this work, a new lattice QCD+QED study has been
reported in Ref.~\cite{Borsanyi:2014jba}. While the results are
compatible with those presented here, it is not clear that the choice
of renormalisation scheme in that work is consistent with the
Cottingham sum rule.


%
\section*{Acknowledgements}
We thank Nathan Hall and James Zanotti for helpful conversations.
This work was supported by the University of Adelaide and the
Australian Research Council through the ARC Centre of Excellence for
Particle Physics at the Terascale and grants FL0992247 (AWT),
DP140103067, FT120100821 (RDY).
\bibliography{HyperonRefs}

\begin{thebibliography}{47}
\expandafter\ifx\csname natexlab\endcsname\relax\def\natexlab#1{#1}\fi
\expandafter\ifx\csname bibnamefont\endcsname\relax
  \def\bibnamefont#1{#1}\fi
\expandafter\ifx\csname bibfnamefont\endcsname\relax
  \def\bibfnamefont#1{#1}\fi
\expandafter\ifx\csname citenamefont\endcsname\relax
  \def\citenamefont#1{#1}\fi
\expandafter\ifx\csname url\endcsname\relax
  \def\url#1{\texttt{#1}}\fi
\expandafter\ifx\csname urlprefix\endcsname\relax\def\urlprefix{URL }\fi
\providecommand{\bibinfo}[2]{#2}
\providecommand{\eprint}[2][]{\url{#2}}

\bibitem[{\citenamefont{Miller et~al.}(2006)\citenamefont{Miller, Opper, and
  Stephenson}}]{Miller:2006tv}
\bibinfo{author}{\bibfnamefont{G.~A.} \bibnamefont{Miller}},
  \bibinfo{author}{\bibfnamefont{A.~K.} \bibnamefont{Opper}}, \bibnamefont{and}
  \bibinfo{author}{\bibfnamefont{E.~J.} \bibnamefont{Stephenson}},
  \bibinfo{journal}{Ann.Rev.Nucl.Part.Sci.} \textbf{\bibinfo{volume}{56}},
  \bibinfo{pages}{253} (\bibinfo{year}{2006}),
\href{http://arxiv.org/abs/nucl-ex/0602021}{{\ttfamily arXiv:nucl-ex/0602021
  [nucl-ex]}}.

\bibitem[{\citenamefont{Londergan et~al.}(2010)\citenamefont{Londergan, Peng,
  and Thomas}}]{Londergan:2009kj}
\bibinfo{author}{\bibfnamefont{J.~T.} \bibnamefont{Londergan}},
  \bibinfo{author}{\bibfnamefont{J.~C.} \bibnamefont{Peng}}, \bibnamefont{and}
  \bibinfo{author}{\bibfnamefont{A.~W.} \bibnamefont{Thomas}},
  \bibinfo{journal}{Rev.Mod.Phys.} \textbf{\bibinfo{volume}{82}},
  \bibinfo{pages}{2009} (\bibinfo{year}{2010}),
\href{http://arxiv.org/abs/0907.2352}{{\ttfamily arXiv:0907.2352 [hep-ph]}}.

\bibitem[{\citenamefont{Gonz\'alez-Alonso and
  Martin-Camalich}(2014)}]{Gonzalez-Alonso:2013ura}
\bibinfo{author}{\bibfnamefont{M.}~\bibnamefont{Gonz\'alez-Alonso}}
  \bibnamefont{and}
  \bibinfo{author}{\bibfnamefont{J.}~\bibnamefont{Martin-Camalich}},
  \bibinfo{journal}{Phys.Rev.Lett.} \textbf{\bibinfo{volume}{112}},
  \bibinfo{pages}{042501} (\bibinfo{year}{2014}),
\href{http://arxiv.org/abs/1309.4434}{{\ttfamily arXiv:1309.4434 [hep-ph]}}.

\bibitem[{\citenamefont{Beane et~al.}(2007)\citenamefont{Beane, Orginos, and
  Savage}}]{Beane:2006fk}
\bibinfo{author}{\bibfnamefont{S.~R.} \bibnamefont{Beane}},
  \bibinfo{author}{\bibfnamefont{K.}~\bibnamefont{Orginos}}, \bibnamefont{and}
  \bibinfo{author}{\bibfnamefont{M.~J.} \bibnamefont{Savage}},
  \bibinfo{journal}{Nucl.Phys.} \textbf{\bibinfo{volume}{B768}},
  \bibinfo{pages}{38} (\bibinfo{year}{2007}),
\href{http://arxiv.org/abs/hep-lat/0605014}{{\ttfamily arXiv:hep-lat/0605014
  [hep-lat]}}.

\bibitem[{\citenamefont{Blum et~al.}(2010)\citenamefont{Blum, Zhou, Doi,
  Hayakawa, Izubuchi et~al.}}]{Blum:2010ym}
\bibinfo{author}{\bibfnamefont{T.}~\bibnamefont{Blum}},
  \bibinfo{author}{\bibfnamefont{R.}~\bibnamefont{Zhou}},
  \bibinfo{author}{\bibfnamefont{T.}~\bibnamefont{Doi}},
  \bibinfo{author}{\bibfnamefont{M.}~\bibnamefont{Hayakawa}},
  \bibinfo{author}{\bibfnamefont{T.}~\bibnamefont{Izubuchi}},
  \bibnamefont{et~al.}, \bibinfo{journal}{Phys.Rev.}
  \textbf{\bibinfo{volume}{D82}}, \bibinfo{pages}{094508}
  (\bibinfo{year}{2010}),
\href{http://arxiv.org/abs/1006.1311}{{\ttfamily arXiv:1006.1311 [hep-lat]}}.

\bibitem[{\citenamefont{de~Divitiis et~al.}(2012)\citenamefont{de~Divitiis,
  Dimopoulos, Frezzotti, Lubicz, Martinelli et~al.}}]{deDivitiis:2011eh}
\bibinfo{author}{\bibfnamefont{G.~M.} \bibnamefont{de~Divitiis}},
  \bibinfo{author}{\bibfnamefont{P.}~\bibnamefont{Dimopoulos}},
  \bibinfo{author}{\bibfnamefont{R.}~\bibnamefont{Frezzotti}},
  \bibinfo{author}{\bibfnamefont{V.}~\bibnamefont{Lubicz}},
  \bibinfo{author}{\bibfnamefont{G.}~\bibnamefont{Martinelli}},
  \bibnamefont{et~al.}, \bibinfo{journal}{JHEP}
  \textbf{\bibinfo{volume}{1204}}, \bibinfo{pages}{124} (\bibinfo{year}{2012}),
\href{http://arxiv.org/abs/1110.6294}{{\ttfamily arXiv:1110.6294 [hep-lat]}}.

\bibitem[{\citenamefont{Horsley et~al.}(2012)}]{Horsley:2012fw}
\bibinfo{author}{\bibfnamefont{R.}~\bibnamefont{Horsley}} \bibnamefont{et~al.}
  (\bibinfo{collaboration}{QCDSF Collaboration, UKQCD Collaboration}),
  \bibinfo{journal}{Phys.Rev.} \textbf{\bibinfo{volume}{D86}},
  \bibinfo{pages}{114511} (\bibinfo{year}{2012}),
\href{http://arxiv.org/abs/1206.3156}{{\ttfamily arXiv:1206.3156 [hep-lat]}}.

\bibitem[{\citenamefont{Shanahan
  et~al.}(2013{\natexlab{a}})\citenamefont{Shanahan, Thomas, and
  Young}}]{Shanahan:2012wa}
\bibinfo{author}{\bibfnamefont{P.~E.} \bibnamefont{Shanahan}},
  \bibinfo{author}{\bibfnamefont{A.~W.} \bibnamefont{Thomas}},
  \bibnamefont{and} \bibinfo{author}{\bibfnamefont{R.~D.} \bibnamefont{Young}},
  \bibinfo{journal}{Phys.Lett.} \textbf{\bibinfo{volume}{B718}},
  \bibinfo{pages}{1148} (\bibinfo{year}{2013}{\natexlab{a}}),
\href{http://arxiv.org/abs/1209.1892}{{\ttfamily arXiv:1209.1892 [nucl-th]}}.

\bibitem[{\citenamefont{Borsanyi et~al.}(2013)\citenamefont{Borsanyi, D{\"u}rr,
  Fodor, Frison, Hoelbling et~al.}}]{Borsanyi:2013lga}
\bibinfo{author}{\bibfnamefont{S.}~\bibnamefont{Borsanyi}},
  \bibinfo{author}{\bibfnamefont{S.}~\bibnamefont{D{\"u}rr}},
  \bibinfo{author}{\bibfnamefont{Z.}~\bibnamefont{Fodor}},
  \bibinfo{author}{\bibfnamefont{J.}~\bibnamefont{Frison}},
  \bibinfo{author}{\bibfnamefont{C.}~\bibnamefont{Hoelbling}},
  \bibnamefont{et~al.}, \bibinfo{journal}{Phys.Rev.Lett.}
  \textbf{\bibinfo{volume}{111}}, \bibinfo{pages}{252001}
  (\bibinfo{year}{2013}),
\href{http://arxiv.org/abs/1306.2287}{{\ttfamily arXiv:1306.2287 [hep-lat]}}.

\bibitem[{\citenamefont{Walker-Loud et~al.}(2012)\citenamefont{Walker-Loud,
  Carlson, and Miller}}]{WalkerLoud:2012bg}
\bibinfo{author}{\bibfnamefont{A.}~\bibnamefont{Walker-Loud}},
  \bibinfo{author}{\bibfnamefont{C.~E.} \bibnamefont{Carlson}},
  \bibnamefont{and} \bibinfo{author}{\bibfnamefont{G.~A.}
  \bibnamefont{Miller}}, \bibinfo{journal}{Phys.Rev.Lett.}
  \textbf{\bibinfo{volume}{108}}, \bibinfo{pages}{232301}
  (\bibinfo{year}{2012}),
\href{http://arxiv.org/abs/1203.0254}{{\ttfamily arXiv:1203.0254 [nucl-th]}}.

\bibitem[{\citenamefont{Cottingham}(1963)}]{Cottingham:1963zz}
\bibinfo{author}{\bibfnamefont{W.~N.} \bibnamefont{Cottingham}},
  \bibinfo{journal}{Annals Phys.} \textbf{\bibinfo{volume}{25}},
  \bibinfo{pages}{424}
 (\bibinfo{year}{1963}).

\bibitem[{\citenamefont{Horsley et~al.}(2013)}]{Horsley:2013qka}
\bibinfo{author}{\bibfnamefont{R.}~\bibnamefont{Horsley}} \bibnamefont{et~al.},
  \bibinfo{journal}{PoS} \textbf{\bibinfo{volume}{Lattice2013}},
  \bibinfo{pages}{499} (\bibinfo{year}{2013}),
\href{http://arxiv.org/abs/1311.4554}{{\ttfamily arXiv:1311.4554 [hep-lat]}}.

\bibitem[{\citenamefont{Gasser and Leutwyler}(1975)}]{Gasser:1974wd}
\bibinfo{author}{\bibfnamefont{J.}~\bibnamefont{Gasser}} \bibnamefont{and}
  \bibinfo{author}{\bibfnamefont{H.}~\bibnamefont{Leutwyler}},
  \bibinfo{journal}{Nucl.Phys.} \textbf{\bibinfo{volume}{B94}},
  \bibinfo{pages}{269}
 (\bibinfo{year}{1975}).

\bibitem[{\citenamefont{Gasser and Leutwyler}(1982)}]{Gasser:1982ap}
\bibinfo{author}{\bibfnamefont{J.}~\bibnamefont{Gasser}} \bibnamefont{and}
  \bibinfo{author}{\bibfnamefont{H.}~\bibnamefont{Leutwyler}},
  \bibinfo{journal}{Phys.Rept.} \textbf{\bibinfo{volume}{87}},
  \bibinfo{pages}{77}
 (\bibinfo{year}{1982}).

\bibitem[{\citenamefont{Thomas et~al.}(2014)\citenamefont{Thomas, Wang, and
  Young}}]{Thomas:2014dxa}
\bibinfo{author}{\bibfnamefont{A.~W.} \bibnamefont{Thomas}},
  \bibinfo{author}{\bibfnamefont{X.~G.} \bibnamefont{Wang}}, \bibnamefont{and}
  \bibinfo{author}{\bibfnamefont{R.~D.} \bibnamefont{Young}}
  (\bibinfo{year}{2014}),
\href{http://arxiv.org/abs/1406.4579}{{\ttfamily arXiv:1406.4579 [nucl-th]}}.

\bibitem[{\citenamefont{Kelly}(2004)}]{Kelly:2004hm}
\bibinfo{author}{\bibfnamefont{J.~J.} \bibnamefont{Kelly}},
  \bibinfo{journal}{Phys.Rev.} \textbf{\bibinfo{volume}{C70}},
  \bibinfo{pages}{068202}
 (\bibinfo{year}{2004}).

\bibitem[{\citenamefont{Shanahan
  et~al.}(2014{\natexlab{a}})}]{Shanahan:2014uka}
\bibinfo{author}{\bibfnamefont{P.~E.} \bibnamefont{Shanahan}}
  \bibnamefont{et~al.}, \bibinfo{journal}{Phys.Rev.}
  \textbf{\bibinfo{volume}{D89}}, \bibinfo{pages}{074511}
  (\bibinfo{year}{2014}{\natexlab{a}}),
\href{http://arxiv.org/abs/1401.5862}{{\ttfamily arXiv:1401.5862 [hep-lat]}}.

\bibitem[{\citenamefont{Shanahan
  et~al.}(2014{\natexlab{b}})}]{Shanahan:2014cga}
\bibinfo{author}{\bibfnamefont{P.~E.} \bibnamefont{Shanahan}}
  \bibnamefont{et~al.}, \bibinfo{journal}{Phys.Rev.}
  \textbf{\bibinfo{volume}{D90}}, \bibinfo{pages}{034502}
  (\bibinfo{year}{2014}{\natexlab{b}}),
\href{http://arxiv.org/abs/1403.1965}{{\ttfamily arXiv:1403.1965 [hep-lat]}}.

\bibitem[{\citenamefont{Beringer et~al.}(2012)}]{Beringer:1900zz}
\bibinfo{author}{\bibfnamefont{J.}~\bibnamefont{Beringer}} \bibnamefont{et~al.}
  (\bibinfo{collaboration}{Particle Data Group}), \bibinfo{journal}{Phys.Rev.}
  \textbf{\bibinfo{volume}{D86}}, \bibinfo{pages}{010001}
 (\bibinfo{year}{2012}).

\bibitem[{\citenamefont{Christy and Bosted}(2010)}]{Christy:2007ve}
\bibinfo{author}{\bibfnamefont{M.~E.} \bibnamefont{Christy}} \bibnamefont{and}
  \bibinfo{author}{\bibfnamefont{P.~E.} \bibnamefont{Bosted}},
  \bibinfo{journal}{Phys.Rev.} \textbf{\bibinfo{volume}{C81}},
  \bibinfo{pages}{055213} (\bibinfo{year}{2010}),
\href{http://arxiv.org/abs/0712.3731}{{\ttfamily arXiv:0712.3731 [hep-ph]}}.

\bibitem[{\citenamefont{Bosted and Christy}(2008)}]{Bosted:2007xd}
\bibinfo{author}{\bibfnamefont{P.~E.} \bibnamefont{Bosted}} \bibnamefont{and}
  \bibinfo{author}{\bibfnamefont{M.~E.} \bibnamefont{Christy}},
  \bibinfo{journal}{Phys.Rev.} \textbf{\bibinfo{volume}{C77}},
  \bibinfo{pages}{065206} (\bibinfo{year}{2008}),
\href{http://arxiv.org/abs/0711.0159}{{\ttfamily arXiv:0711.0159 [hep-ph]}}.

\bibitem[{\citenamefont{Capella et~al.}(1994)\citenamefont{Capella, Kaidalov,
  Merino, and Tran Thanh~Van}}]{Capella:1994cr}
\bibinfo{author}{\bibfnamefont{A.}~\bibnamefont{Capella}},
  \bibinfo{author}{\bibfnamefont{A.}~\bibnamefont{Kaidalov}},
  \bibinfo{author}{\bibfnamefont{C.}~\bibnamefont{Merino}}, \bibnamefont{and}
  \bibinfo{author}{\bibfnamefont{J.}~\bibnamefont{Tran Thanh~Van}},
  \bibinfo{journal}{Phys.Lett.} \textbf{\bibinfo{volume}{B337}},
  \bibinfo{pages}{358} (\bibinfo{year}{1994}),
\href{http://arxiv.org/abs/hep-ph/9405338}{{\ttfamily arXiv:hep-ph/9405338
  [hep-ph]}}.

\bibitem[{\citenamefont{Sibirtsev et~al.}(2010)\citenamefont{Sibirtsev,
  Blunden, Melnitchouk, and Thomas}}]{Sibirtsev:2010zg}
\bibinfo{author}{\bibfnamefont{A.}~\bibnamefont{Sibirtsev}},
  \bibinfo{author}{\bibfnamefont{P.~G.} \bibnamefont{Blunden}},
  \bibinfo{author}{\bibfnamefont{W.}~\bibnamefont{Melnitchouk}},
  \bibnamefont{and} \bibinfo{author}{\bibfnamefont{A.~W.}
  \bibnamefont{Thomas}}, \bibinfo{journal}{Phys.Rev.}
  \textbf{\bibinfo{volume}{D82}}, \bibinfo{pages}{013011}
  (\bibinfo{year}{2010}),
\href{http://arxiv.org/abs/1002.0740}{{\ttfamily arXiv:1002.0740 [hep-ph]}}.

\bibitem[{\citenamefont{Boros and Thomas}(1999)}]{Boros:1999tb}
\bibinfo{author}{\bibfnamefont{C.}~\bibnamefont{Boros}} \bibnamefont{and}
  \bibinfo{author}{\bibfnamefont{A.~W.} \bibnamefont{Thomas}},
  \bibinfo{journal}{Phys.Rev.} \textbf{\bibinfo{volume}{D60}},
  \bibinfo{pages}{074017} (\bibinfo{year}{1999}),
\href{http://arxiv.org/abs/hep-ph/9902372}{{\ttfamily arXiv:hep-ph/9902372
  [hep-ph]}}.

\bibitem[{\citenamefont{Horsley et~al.}(2011)}]{Horsley:2010th}
\bibinfo{author}{\bibfnamefont{R.}~\bibnamefont{Horsley}} \bibnamefont{et~al.},
  \bibinfo{journal}{Phys.Rev.} \textbf{\bibinfo{volume}{D83}},
  \bibinfo{pages}{051501} (\bibinfo{year}{2011}),
\href{http://arxiv.org/abs/1012.0215}{{\ttfamily arXiv:1012.0215 [hep-lat]}}.

\bibitem[{\citenamefont{Clo\"et et~al.}(2012)\citenamefont{Clo\"et, Horsley,
  Londergan, Nakamura, Pleiter et~al.}}]{Cloet:2012db}
\bibinfo{author}{\bibfnamefont{I.~C.} \bibnamefont{Clo\"et}},
  \bibinfo{author}{\bibfnamefont{R.}~\bibnamefont{Horsley}},
  \bibinfo{author}{\bibfnamefont{J.~T.} \bibnamefont{Londergan}},
  \bibinfo{author}{\bibfnamefont{Y.}~\bibnamefont{Nakamura}},
  \bibinfo{author}{\bibfnamefont{D.}~\bibnamefont{Pleiter}},
  \bibnamefont{et~al.}, \bibinfo{journal}{Phys.Lett.}
  \textbf{\bibinfo{volume}{B714}}, \bibinfo{pages}{97} (\bibinfo{year}{2012}),
\href{http://arxiv.org/abs/1204.3492}{{\ttfamily arXiv:1204.3492 [hep-lat]}}.

\bibitem[{\citenamefont{Shanahan
  et~al.}(2013{\natexlab{b}})\citenamefont{Shanahan, Thomas, and
  Young}}]{Shanahan:2013vla}
\bibinfo{author}{\bibfnamefont{P.~E.} \bibnamefont{Shanahan}},
  \bibinfo{author}{\bibfnamefont{A.~W.} \bibnamefont{Thomas}},
  \bibnamefont{and} \bibinfo{author}{\bibfnamefont{R.~D.} \bibnamefont{Young}},
  \bibinfo{journal}{Phys. Rev. D 87,} \textbf{\bibinfo{volume}{094515}}
  (\bibinfo{year}{2013}{\natexlab{b}}),
\href{http://arxiv.org/abs/1303.4806}{{\ttfamily arXiv:1303.4806 [nucl-th]}}.

\bibitem[{\citenamefont{Melnitchouk et~al.}(2005)\citenamefont{Melnitchouk,
  Ent, and Keppel}}]{Melnitchouk:2005zr}
\bibinfo{author}{\bibfnamefont{W.}~\bibnamefont{Melnitchouk}},
  \bibinfo{author}{\bibfnamefont{R.}~\bibnamefont{Ent}}, \bibnamefont{and}
  \bibinfo{author}{\bibfnamefont{C.}~\bibnamefont{Keppel}},
  \bibinfo{journal}{Phys.Rept.} \textbf{\bibinfo{volume}{406}},
  \bibinfo{pages}{127} (\bibinfo{year}{2005}),
\href{http://arxiv.org/abs/hep-ph/0501217}{{\ttfamily arXiv:hep-ph/0501217
  [hep-ph]}}.

\bibitem[{\citenamefont{Rislow and Carlson}(2011)}]{Rislow:2010vi}
\bibinfo{author}{\bibfnamefont{B.~C.} \bibnamefont{Rislow}} \bibnamefont{and}
  \bibinfo{author}{\bibfnamefont{C.~E.} \bibnamefont{Carlson}},
  \bibinfo{journal}{Phys.Rev.} \textbf{\bibinfo{volume}{D83}},
  \bibinfo{pages}{113007} (\bibinfo{year}{2011}),
\href{http://arxiv.org/abs/1011.2397}{{\ttfamily arXiv:1011.2397 [hep-ph]}}.

\bibitem[{\citenamefont{Gorchtein et~al.}(2011)\citenamefont{Gorchtein,
  Horowitz, and Ramsey-Musolf}}]{Gorchtein:2011mz}
\bibinfo{author}{\bibfnamefont{M.}~\bibnamefont{Gorchtein}},
  \bibinfo{author}{\bibfnamefont{C.~J.} \bibnamefont{Horowitz}},
  \bibnamefont{and} \bibinfo{author}{\bibfnamefont{M.~J.}
  \bibnamefont{Ramsey-Musolf}}, \bibinfo{journal}{Phys.Rev.}
  \textbf{\bibinfo{volume}{C84}}, \bibinfo{pages}{015502}
  (\bibinfo{year}{2011}),
\href{http://arxiv.org/abs/1102.3910}{{\ttfamily arXiv:1102.3910 [nucl-th]}}.

\bibitem[{\citenamefont{Hall et~al.}(2013)\citenamefont{Hall, Blunden,
  Melnitchouk, Thomas, and Young}}]{Hall:2013hta}
\bibinfo{author}{\bibfnamefont{N.~L.} \bibnamefont{Hall}},
  \bibinfo{author}{\bibfnamefont{P.~G.} \bibnamefont{Blunden}},
  \bibinfo{author}{\bibfnamefont{W.}~\bibnamefont{Melnitchouk}},
  \bibinfo{author}{\bibfnamefont{A.~W.} \bibnamefont{Thomas}},
  \bibnamefont{and} \bibinfo{author}{\bibfnamefont{R.~D.} \bibnamefont{Young}},
  \bibinfo{journal}{Phys.Rev.} \textbf{\bibinfo{volume}{D88}},
  \bibinfo{pages}{013011} (\bibinfo{year}{2013}),
\href{http://arxiv.org/abs/1304.7877}{{\ttfamily arXiv:1304.7877 [nucl-th]}}.

\bibitem[{\citenamefont{Carlson and Vanderhaeghen}(2011)}]{Carlson:2011zd}
\bibinfo{author}{\bibfnamefont{C.~E.} \bibnamefont{Carlson}} \bibnamefont{and}
  \bibinfo{author}{\bibfnamefont{M.}~\bibnamefont{Vanderhaeghen}},
  \bibinfo{journal}{Phys.Rev.} \textbf{\bibinfo{volume}{A84}},
  \bibinfo{pages}{020102} (\bibinfo{year}{2011}),
\href{http://arxiv.org/abs/1101.5965}{{\ttfamily arXiv:1101.5965 [hep-ph]}}.

\bibitem[{\citenamefont{Hill and Paz}(2011)}]{Hill:2011wy}
\bibinfo{author}{\bibfnamefont{R.~J.} \bibnamefont{Hill}} \bibnamefont{and}
  \bibinfo{author}{\bibfnamefont{G.}~\bibnamefont{Paz}},
  \bibinfo{journal}{Phys.Rev.Lett.} \textbf{\bibinfo{volume}{107}},
  \bibinfo{pages}{160402} (\bibinfo{year}{2011}),
\href{http://arxiv.org/abs/1103.4617}{{\ttfamily arXiv:1103.4617 [hep-ph]}}.

\bibitem[{\citenamefont{Birse and McGovern}(2012)}]{Birse:2012eb}
\bibinfo{author}{\bibfnamefont{M.~C.} \bibnamefont{Birse}} \bibnamefont{and}
  \bibinfo{author}{\bibfnamefont{J.~A.} \bibnamefont{McGovern}},
  \bibinfo{journal}{Eur.Phys.J.} \textbf{\bibinfo{volume}{A48}},
  \bibinfo{pages}{120} (\bibinfo{year}{2012}),
\href{http://arxiv.org/abs/1206.3030}{{\ttfamily arXiv:1206.3030 [hep-ph]}}.

\bibitem[{\citenamefont{Pohl et~al.}(2010)}]{Pohl:2010zza}
\bibinfo{author}{\bibfnamefont{R.}~\bibnamefont{Pohl}} \bibnamefont{et~al.},
  \bibinfo{journal}{Nature} \textbf{\bibinfo{volume}{466}},
  \bibinfo{pages}{213}
 (\bibinfo{year}{2010}).

\bibitem[{\citenamefont{Pohl et~al.}(2013)\citenamefont{Pohl, Gilman, Miller,
  and Pachucki}}]{Pohl:2013yb}
\bibinfo{author}{\bibfnamefont{R.}~\bibnamefont{Pohl}},
  \bibinfo{author}{\bibfnamefont{R.}~\bibnamefont{Gilman}},
  \bibinfo{author}{\bibfnamefont{G.~A.} \bibnamefont{Miller}},
  \bibnamefont{and} \bibinfo{author}{\bibfnamefont{K.}~\bibnamefont{Pachucki}},
  \bibinfo{journal}{Ann.Rev.Nucl.Part.Sci.} \textbf{\bibinfo{volume}{63}},
  \bibinfo{pages}{175} (\bibinfo{year}{2013}),
\href{http://arxiv.org/abs/1301.0905}{{\ttfamily arXiv:1301.0905
  [physics.atom-ph]}}.

\bibitem[{\citenamefont{Bernabeu and Tarrach}(1976)}]{Bernabeu:1976jq}
\bibinfo{author}{\bibfnamefont{J.}~\bibnamefont{Bernabeu}} \bibnamefont{and}
  \bibinfo{author}{\bibfnamefont{R.}~\bibnamefont{Tarrach}},
  \bibinfo{journal}{Annals Phys.} \textbf{\bibinfo{volume}{102}},
  \bibinfo{pages}{323}
 (\bibinfo{year}{1976}).

\bibitem[{\citenamefont{Grie\ss{}hammer
  et~al.}(2012)\citenamefont{Grie\ss{}hammer, McGovern, Phillips, and
  Feldman}}]{Griesshammer:2012we}
\bibinfo{author}{\bibfnamefont{H.~W.} \bibnamefont{Grie\ss{}hammer}},
  \bibinfo{author}{\bibfnamefont{J.~A.} \bibnamefont{McGovern}},
  \bibinfo{author}{\bibfnamefont{D.~R.} \bibnamefont{Phillips}},
  \bibnamefont{and} \bibinfo{author}{\bibfnamefont{G.}~\bibnamefont{Feldman}},
  \bibinfo{journal}{Prog.Part.Nucl.Phys.} \textbf{\bibinfo{volume}{67}},
  \bibinfo{pages}{841} (\bibinfo{year}{2012}),
\href{http://arxiv.org/abs/1203.6834}{{\ttfamily arXiv:1203.6834 [nucl-th]}}.

\bibitem[{\citenamefont{Collins}(1979)}]{Collins:1978hi}
\bibinfo{author}{\bibfnamefont{J.~C.} \bibnamefont{Collins}},
  \bibinfo{journal}{Nucl.Phys.} \textbf{\bibinfo{volume}{B149}},
  \bibinfo{pages}{90}
 (\bibinfo{year}{1979}).

\bibitem[{\citenamefont{Shanahan et~al.}(2012)\citenamefont{Shanahan, Thomas,
  and Young}}]{Shanahan:2013cd}
\bibinfo{author}{\bibfnamefont{P.~E.} \bibnamefont{Shanahan}},
  \bibinfo{author}{\bibfnamefont{A.~W.} \bibnamefont{Thomas}},
  \bibnamefont{and} \bibinfo{author}{\bibfnamefont{R.~D.} \bibnamefont{Young}},
  \bibinfo{journal}{PoS} \textbf{\bibinfo{volume}{LATTICE2012}},
  \bibinfo{pages}{165} (\bibinfo{year}{2012}),
\href{http://arxiv.org/abs/1301.3231}{{\ttfamily arXiv:1301.3231 [hep-lat]}}.

\bibitem[{\citenamefont{Shanahan
  et~al.}(2013{\natexlab{c}})\citenamefont{Shanahan, Thomas, and
  Young}}]{Shanahan:2012wh}
\bibinfo{author}{\bibfnamefont{P.~E.} \bibnamefont{Shanahan}},
  \bibinfo{author}{\bibfnamefont{A.~W.} \bibnamefont{Thomas}},
  \bibnamefont{and} \bibinfo{author}{\bibfnamefont{R.~D.} \bibnamefont{Young}},
  \bibinfo{journal}{Phys.Rev.} \textbf{\bibinfo{volume}{D87}},
  \bibinfo{pages}{074503} (\bibinfo{year}{2013}{\natexlab{c}}),
\href{http://arxiv.org/abs/1205.5365}{{\ttfamily arXiv:1205.5365 [nucl-th]}}.

\bibitem[{\citenamefont{Bernard et~al.}(1992)\citenamefont{Bernard, Kaiser,
  Kambor, and Mei\ss{}ner}}]{Bernard:1992xi}
\bibinfo{author}{\bibfnamefont{V.}~\bibnamefont{Bernard}},
  \bibinfo{author}{\bibfnamefont{N.}~\bibnamefont{Kaiser}},
  \bibinfo{author}{\bibfnamefont{J.}~\bibnamefont{Kambor}}, \bibnamefont{and}
  \bibinfo{author}{\bibfnamefont{U.-G.} \bibnamefont{Mei\ss{}ner}},
  \bibinfo{journal}{Phys.Rev.} \textbf{\bibinfo{volume}{D46}},
  \bibinfo{pages}{2756}
 (\bibinfo{year}{1992}).

\bibitem[{\citenamefont{Gobbi et~al.}(1996)\citenamefont{Gobbi, Schat, and
  Scoccola}}]{Gobbi:1995de}
\bibinfo{author}{\bibfnamefont{C.}~\bibnamefont{Gobbi}},
  \bibinfo{author}{\bibfnamefont{C.~L.} \bibnamefont{Schat}}, \bibnamefont{and}
  \bibinfo{author}{\bibfnamefont{N.~N.} \bibnamefont{Scoccola}},
  \bibinfo{journal}{Nucl.Phys.} \textbf{\bibinfo{volume}{A598}},
  \bibinfo{pages}{318} (\bibinfo{year}{1996}),
\href{http://arxiv.org/abs/hep-ph/9509211}{{\ttfamily arXiv:hep-ph/9509211
  [hep-ph]}}.

\bibitem[{\citenamefont{Tanushi et~al.}(2000)\citenamefont{Tanushi, Saito, and
  Uehara}}]{Tanushi:1999rq}
\bibinfo{author}{\bibfnamefont{Y.}~\bibnamefont{Tanushi}},
  \bibinfo{author}{\bibfnamefont{S.}~\bibnamefont{Saito}}, \bibnamefont{and}
  \bibinfo{author}{\bibfnamefont{M.}~\bibnamefont{Uehara}},
  \bibinfo{journal}{Phys.Rev.} \textbf{\bibinfo{volume}{C61}},
  \bibinfo{pages}{055204} (\bibinfo{year}{2000}),
\href{http://arxiv.org/abs/nucl-th/9911071}{{\ttfamily arXiv:nucl-th/9911071
  [nucl-th]}}.

\bibitem[{\citenamefont{Aleksejevs and Barkanova}(2011)}]{Aleksejevs:2010zw}
\bibinfo{author}{\bibfnamefont{A.}~\bibnamefont{Aleksejevs}} \bibnamefont{and}
  \bibinfo{author}{\bibfnamefont{S.}~\bibnamefont{Barkanova}},
  \bibinfo{journal}{J.Phys.} \textbf{\bibinfo{volume}{G38}},
  \bibinfo{pages}{035004} (\bibinfo{year}{2011}),
\href{http://arxiv.org/abs/1010.3457}{{\ttfamily arXiv:1010.3457 [nucl-th]}}.

\bibitem[{\citenamefont{Lee et~al.}(2006)\citenamefont{Lee, Zhou, Wilcox, and
  Christensen}}]{Lee:2005dq}
\bibinfo{author}{\bibfnamefont{F.~X.} \bibnamefont{Lee}},
  \bibinfo{author}{\bibfnamefont{L.}~\bibnamefont{Zhou}},
  \bibinfo{author}{\bibfnamefont{W.}~\bibnamefont{Wilcox}}, \bibnamefont{and}
  \bibinfo{author}{\bibfnamefont{J.~C.} \bibnamefont{Christensen}},
  \bibinfo{journal}{Phys.Rev.} \textbf{\bibinfo{volume}{D73}},
  \bibinfo{pages}{034503} (\bibinfo{year}{2006}),
\href{http://arxiv.org/abs/hep-lat/0509065}{{\ttfamily arXiv:hep-lat/0509065
  [hep-lat]}}.

\bibitem[{\citenamefont{Borsanyi et~al.}(2014)\citenamefont{Borsanyi, Durr,
  Fodor, Hoelbling, Katz et~al.}}]{Borsanyi:2014jba}
\bibinfo{author}{\bibfnamefont{S.}~\bibnamefont{Borsanyi}},
  \bibinfo{author}{\bibfnamefont{S.}~\bibnamefont{Durr}},
  \bibinfo{author}{\bibfnamefont{Z.}~\bibnamefont{Fodor}},
  \bibinfo{author}{\bibfnamefont{C.}~\bibnamefont{Hoelbling}},
  \bibinfo{author}{\bibfnamefont{S.}~\bibnamefont{Katz}}, \bibnamefont{et~al.}
  (\bibinfo{year}{2014}),
\href{http://arxiv.org/abs/1406.4088}{{\ttfamily arXiv:1406.4088 [hep-lat]}}.

\end{thebibliography}

\end{document}